\documentclass[11pt]{article}

\usepackage{amsfonts}
\usepackage{amssymb}
\usepackage{amsmath}
\usepackage{amsthm}
\usepackage{amsfonts}
\usepackage{multicol}
\usepackage{amscd}
\usepackage{textcomp}
\usepackage{multirow}
\usepackage{times}
\usepackage{color}
\usepackage[english, activeacute]{babel}

\def\be{\begin{equation}}
\def\ee{\end{equation}}
\def\bea{\begin{eqnarray}}
\def\eea{\end{eqnarray}}

\def\pois#1#2{\left\{ {#1},{#2} \right\}}

\def\1{\'{\i}}

\def\>#1{{\mathbf#1}}

\begin{document}

\thispagestyle{empty}


\ 
\vspace{0.5cm}

\begin{center}

{\LARGE{\sc{Quantum algebras as quantizations of dual 
Poisson-Lie groups
}} }

\end{center}

\begin{center} {\sc \'Angel Ballesteros$^a$ and Fabio Musso$^b$}
\end{center}

\begin{center} {$^a$\it{Departamento de F\1sica,  Universidad de Burgos, 
09001 Burgos, Spain}}

{$^b$\it{Dipartimento di Fisica `Edoardo Amaldi',  Universit\'a degli Studi di Roma Tre, 
00146 Roma, Italy}}

e-mail: angelb@ubu.es, musso@fis.uniroma3.it
\end{center}

  \medskip

\begin{abstract} 
\noindent 

A systematic computational approach for the explicit construction of any quantum Hopf algebra $(U_z(g),\Delta_z)$ starting from the Lie bialgebra $(g,\delta)$ that gives the first-order deformation of the coproduct map $\Delta_z$ is presented. The procedure is based on the well-known `quantum duality principle', namely, the fact that any quantum algebra can be viewed as the quantization of the unique Poisson-Lie structure $(G^\ast,\Lambda_g)$ on the dual group $G^\ast$, which is obtained by exponentiating the Lie algebra $g^\ast$ defined by the dual map $\delta^\ast$. From this perspective, the coproduct for $U_z(g)$ is just the pullback of the group law for $G^\ast$, and the Poisson analogues of the quantum commutation rules for $U_z(g)$ are given by the unique Poisson-Lie structure $\Lambda_g$ on $G^\ast$ whose linearization is the Poisson analogue of the initial Lie algebra $g$. This approach is shown to be a very useful technical tool in order to solve the Lie bialgebra quantization problem explicitly since, once a Lie bialgebra $(g,\delta)$ is given, the full dual Poisson-Lie group $(G^\ast,\Lambda)$ can be obtained either by applying standard Poisson-Lie group techniques or by implementing the algorithm here presented with the aid of symbolic manipulation programs. As a consequence, the quantization of $(G^\ast,\Lambda)$ will give rise to the full $U_z(g)$ quantum algebra, provided that ordering problems are appropriately fixed through the choice of certain local coordinates on $G^\ast$ whose coproduct fulfills a precise `quantum symmetry' property. The applicability of this approach is explicitly demonstrated by  reviewing the construction of several  instances of quantum deformations of physically relevant Lie algebras as $sl(2,\mathbb{R})$, the (2+1) Anti de Sitter algebra $so(2,2)$ and the Poincar\'e algebra in (3+1) dimensions.
 \end{abstract}

\bigskip\bigskip\bigskip\bigskip



\vfill
\newpage


\section{Introduction}

Quantum algebras $(U_z(g),\Delta_z)$ are Hopf algebra deformations of universal enveloping algebras $U(g)$  `in the direction' of a given Lie bialgebra $(g,\delta)$ (we refer the reader to~\cite{Dri}-\cite{majid} for all basic notions in quantum group theory). This means that given a compatible quantum coproduct $\Delta_z$ for $U_z(g)$, a
unique Lie bialgebra structure
$(g,\delta)$ arises as the skew-symmetric part of the first-order of $\Delta_z$ in terms of the deformation parameter(s) $z$, namely,
\be
\delta (X)=\frac 1 2 (\Delta_z (X)-\sigma\circ \Delta_z (X)) ,\quad
\mbox{mod}\,z^2 ,\qquad\forall X\in g ,
\ee
where $ \sigma$ is the flip operator  $ \sigma (X\otimes
Y)=Y\otimes X$. Such skew-symmetric map $\delta:g\rightarrow g\otimes g$ is the so-called cocommutator map. Moreover, we recall that Lie bialgebras $(g,\delta)$
are in one to one correspondence with Poisson-Lie (hereafter PL) structures on the
(simply connected) group $G\equiv \mbox{Exp}(g)$ via Drinfel'd theorem~\cite{DriPL}, and the quantization of the unique PL structure associated to $(g,\delta)$ will be the quantum group that is dual (in the Hopf algebra sense) to the quantum algebra
$(U_z(g),\Delta_z)$~\cite{Semquantum,Bidegain}.

Therefore, quantum deformations of a given 
$U(g)$ can be identified according to their
`semi-classical' limit given by the Lie bialgebras $(g,\delta)$. With this in mind, classifications of Lie bialgebra structures for
several physically relevant Lie algebras have been obtained. For the three
dimensional case  we recall the complete classification given in \cite{gomez} and for some higher dimensional Lie bialgebras we refer to
\cite{BBM3d} and references therein. Let us also stress that Lie bialgebras and PL groups are interesting on their own in the theory of classical integrable systems \cite{DriPL} and they also appear as Poisson manifolds for some interesting Hamiltonian systems (see, for example, \cite{LV}).

Due to the relevance of quantum algebras in very different mathematical and physical contexts (see~\cite{Dri}-\cite{majid}), it becomes clear that the generic  `Lie bialgebra quantization' problem (see~\cite{EK,EH}) is worth to be faced: this means that, given a Lie bialgebra
$(g,\delta)$, one should obtain the most general recipes for the `quantization' (i.e.~the construction of the all-orders quantum universal enveloping algebra) whose first order deformation is given by $(g,\delta)$. Although the existence of such quantization is indeed guaranteed (see
\cite{charipressley}, Chapter 6), only for coboundary triangular
Lie bialgebras the Drinfel'd twist
operator~\cite{Drtwist} gives the answer
(see, for instance, \cite{Reshetikhin,Mudrov,twistedP,KLM,LukiWoroPLB,Dasz}). Nevertheless, no generic way to express the twist operator in terms of $\delta$ is known. Moreover, quasitriangular and non-coboundary deformations do exist, and for them
the twist operator approach to quantization is not available. Therefore, it seems that the only completely general way to proceed is by using an order-by-order quantization in which higher orders of the deformation are obtained from previous ones by imposing both the coassociativity condition on the coproduct and, simultaneously, the homomorphism condition with respect to the deformed commutation rules (see, for instance,~\cite{Tjin,ballesteros04}).

As a consequence, it would be certainly interesting to develop a general method in order to `exponentiate' the Lie bialgebra $(g,\delta)$ in order to get the full quantum algebra $(U_z(g),\Delta_z)$. In this paper we present a general procedure to do this that, moreover, is amenable to be implemented to a high extent by making use of symbolic manipulation packages. Essentially, the approach we propose is  to make use of the well-known `quantum duality principle' (see~\cite{Dri, STS, Gav, GC} and references therein) that asserts that a quantum universal enveloping algebra can be always thought of as a quantum dual group. The foundations of this principle can be summarized as follows:

\begin{itemize}

\item  Given a Lie bialgebra $(g,\delta)$ we know that the dual Lie bialgebra $(g^\ast,\delta^\ast)$ can be always constructed immediately through the canonical pairing connecting $g$ and $g^\ast$ since, by definition, the map $\delta$ provides a Lie algebra structure on $g^\ast$, and the Lie algebra structure on $g$ 
gives rise to the dual cocommutator map
$\delta^\ast: g^\ast\rightarrow g^\ast\otimes g^\ast$.

\item If we succeed in quantizing both Lie bialgebras $(g,\delta)$  and $(g^\ast,\delta^\ast)$, we would obtain two different quantum universal enveloping algebras $(U_z(g),\Delta_z)$ and $(U_z(g^\ast),\tilde\Delta_z)$ that, by construction, will be dual Hopf algebras in the sense that we can define a canonical pairing
\be
\langle
\, , \, \rangle: U_z(g^\ast)\times U_z(g)\rightarrow \mathbb C,
\ee
such that
\bea
&& \langle m_{U_z(g^\ast)}(f\otimes g), a \rangle=\langle f\otimes g ,
\Delta_{U_z(g)}(a)\rangle,
\\
&& \langle \Delta_{U_z(g^\ast)}(f) , a\otimes b \rangle= \langle f,
m_{U_z(g)}(a\otimes b)\rangle,
\eea
where $a,b\in U_z(g)$; $f,g\in U_z(g^\ast)$; $m$ is the product map for the corresponding quantum algebra and 
$
\langle f\otimes g , a\otimes b \rangle=\langle f , a
\rangle\,\langle g , b \rangle.
$
In fact, after the Hopf algebra quantization is performed, these relations extend the first order duality between $(g,\delta)$ and $(g^\ast,\delta^\ast)$ to all orders in the deformation parameter. 

\item Also, we know that $(g^\ast,\delta^\ast)$ is the infinitesimal counterpart of a unique PL structure $(G^\ast,\Lambda)$ on the group $G^\ast\equiv\mbox{Exp}(g^\ast)$, where $\Lambda$ is the corresponding Poisson bivector. This a Poisson-Hopf algebra $(G^\ast,\Lambda,\Delta_{G^\ast})$ whose coproduct $\Delta_{G^\ast}$ will be given by the group multiplication on $G^\ast$ (more technically, by its pull-back, as detailed in~\cite{BBMpullback}).

\item Therefore, the quantization of the latter PL structure (as a Hopf algebra) will give rise to the quantum group $(\hat G^\ast,\Delta_{G^\ast})$ which is, again by construction, the Hopf algebra dual to the quantum universal enveloping algebra $(U_z(g^\ast),\tilde\Delta_z)$. 

\item Finally, since $(U_z(g),\Delta_z)$ is also the Hopf algebra dual of $(U_z(g^\ast),\tilde\Delta_z)$, then the former Hopf algebra can be identified with $(\hat G^\ast,\Delta)$, which turns out to be a Hopf algebra quantization of $(g,\delta)$. 

\end{itemize}

The aim of this paper is to show that this constructive way of presenting the `quantum duality principle' can be helpful in order to overcome many technical difficulties that appear when the quantization problem for a given Lie bialgebra has to be explicitly solved.

As we will show in the sequel,  
the main advantage of this dual PL group approach to quantization stems from the fact that nowadays the explicit construction of PL structures is affordable by using symbolic computation programs (see~\cite{BBM3d}). 
Namely, given a Lie bialgebra $(g,\delta)$, we will see that the Poisson bracket  $\Lambda$ on the dual group (which is the classical counterpart of the $U_z(g)$ commutation rules) can be explicitly obtained even for high dimensional Lie bialgebras. Moreover, the compatible coproduct $\Delta_{G^\ast}$ will be given by the pull-back of the group multiplication law for  $G^\ast$, that always exists and can be explicitly obtained, even in quite complicated cases. Once all this classical information is known, the only task to be performed by hand would be the quantization of the (commutative, non-cocommutative) Poisson Hopf algebra $(G^\ast,\Lambda,\Delta_{G^\ast})$ into the (non-commutative, non-cocommutative) Hopf algebra $(\hat G^\ast,\Delta_z)\equiv (U_z(g),\Delta_z)$. This `standard' quantization process is straightforward if we are able to find some local coordinates on $G^\ast$ for which the Poisson bracket $\Lambda$ and the group multiplication $\Delta_{G^\ast}$ minimize the number of ordering ambiguities. Again, to find these coordinates is a problem defined on commutative variables and, as we shall see, it can be solved in all the examples here presented in terms of certain local coordinates for $G^\ast$ that fulfill certain `quantum symmetry' property. Nevertheless, we stress that even in the case that this set of coordinates would not be apparent, the quantization of $(G^\ast,\Lambda,\Delta_{G^\ast})$ could be always obtained by using generic (although more cumbersome) symmetrization procedures (see~\cite{ballesteros04}). 

To the best of our knowledge, the specific procedure described in the previous paragraph has not been fully exploited so far, since only the rather elementary example of the Heisenberg Lie algebra was presented in detail in~\cite{BHOSpl}. Also, it is worth to recall that this approach underlies the quantization prescription for certain classes of Lie bialgebras with solvable dual Lie group $G^\ast$ that was proposed in~\cite{LM,MudrovPet,MudrovJMP}. Therefore, in this paper we review the explicit construction of several known quantizations of Lie bialgebras (with different representative dimensions) through the previuosly described implementation of the `quantum duality principle', thus showing that this technique could be helpful in order to construct new quantum deformations of interest.

The paper is organized as follows. In the next section the algorithm is presented in full detail, and is illustrated with the construction step by step of the Drinfel'd-Jimbo~\cite{Dri,Jimbo} deformation for $sl(2,\mathbb{R})$
as the quantization of a PL structure on the corresponding dual group (which is the so-called `book' Lie group). Section 3 is devoted to show a first non-trivial case from the computational viewpoint: The Drinfel'd-Jimbo quantization 
for the Anti de Sitter algebra $so(2,2)$. In Section 4 the well-kown $\kappa$-deformation of Poincar\'e algebra in (3+1) dimensions~\cite{LNRT,LNR} is recovered and, in order to show the potentialities of this method, a twisted $\kappa$-Poincar\'e algebra~\cite{Dasz} is further obtained in Section 5. The paper is closed by a final Section that includes some remarks and open problems.


\section{The basic example: quantum $sl(2,\mathbb{R})$}
 
In order to make the construction more transparent, in this Section we will present the algorithm through the most basic and representative three-dimensional example: the Drinfel'd-Jimbo quantum deformation of $sl(2,\mathbb{R})$. We will split the procedure in several specific steps, and we will fix the notation that is going to be used throughout the paper. We remark that we will be always concerned with real Lie algebras and groups, therefore all parameters have to be real ones.

\subsection{The initial data: $(g,\delta)$}

The quantum universal enveloping algebra $U_z(g)$ that we want to obtain is defined through the `direction' of the deformation given by the chosen Lie bialgebra structure ${(g,\delta)}$. We put ${\rm dim} (g)=d$, and the Lie bialgebra has to be explicitly given by the structure tensors $c_{ij}^k$ and $f_{i}^{jk}$ in the form
\be
\left[ x_i, x_j \right] =c_{ij}^k\, x_k 
\qquad 
\delta(x_i)= f_{i}^{jk} x_j \wedge x_k 
\ee
where $f_{i}^{jk}$ are linear functions of the quantum deformation parameter(s).

In the $sl(2,\mathbb{R})$ case, we consider the Lie algebra generators $\{j_3,j_+,j_-\}$ fulfilling
\begin{equation}
[j_3,j_\pm]=\pm 2 j_\pm, \quad [j_+,j_-]=j_3.
\end{equation}
In the case of coboundary Lie bialgebras, the cocommutator $\delta$ will be generated by a classical $r-$matrix through 
\begin{equation}
\delta(x)=[1\otimes x + x\otimes 1, r],\qquad \forall x\in g.
\label{rmatrix}
\end{equation}
For the Drinfel'd-Jimbo deformation of $sl(2,\mathbb{R})$ we are looking for, we have that
\be
r=z \, J_+ \wedge J_-,
\ee
from which we obtain
\be
\delta(j_3)=0 \qquad
\delta(j_+)= z \, j_+ \wedge j_3\qquad
\delta(j_-)= z \, j_- \wedge j_3 .
\ee
Note that we will always include the deformation parameter(s) within the structure tensor $f_{i}^{jk}$.

\subsection{The dual Lie bialgebra $(g^\ast,\delta^\ast)$}

The dual structure $(g^\ast,\delta^\ast)$ with  basis $\{x^1,\dots,x^d\}$ is constructed by interchanging the structure tensors, namely:
\be
[ x^j, x^k ]=  f^{jk}_i x^i
\qquad 
\delta^\ast(x^k)=c_{ij}^k\,  x^i \wedge x^j. 
\ee

In the the $sl(2,\mathbb{R})$ case, let $\{j^3,j^+,j^-\}$ be such a basis. Then the dual Lie bialgebra $(g^\ast,\delta^\ast)$ is given by:
\begin{equation}
[j^3,j^\pm]=-z \, j^\pm \qquad [j^+,j^-]=0
\label{book}
\end{equation}
\begin{equation}
\delta^\ast(j^3)=j_+\wedge j^- \qquad
\delta^\ast(j^+)= 2\, j^3\wedge j^+\qquad
\delta^\ast(j^-)= -2\, j^3\wedge j^- .
\end{equation}
The solvable Lie algebra (\ref{book}) is the so-called `book Lie algebra' (see~\cite{BHOSpl} and references therein). In order to construct the Poisson-Lie group associated with $(g^\ast,\delta^\ast)$ it is important to know whether this is a coboundary Lie bialgebra, i.e., to check if there exists a classical $r$-matrix 
\be
r^\ast=\alpha\, j^3\wedge j^+ + \beta\, j^3\wedge j^- + \gamma\, j^+\wedge j^-
\ee
such that
\be
\delta^\ast(x)=[1\otimes x + x\otimes 1,r^\ast], \qquad \forall x\in g^\ast.
\ee
It is straightforward to check that in this case there is no solution for $\alpha,\beta,\gamma$, and $(g^\ast,\delta^\ast)$ is a non-coboundary Lie bialgebra (see~\cite{gomez,BHOSpl} for the full classification of real three dimensional Lie bialgebras). 

\subsection{Construction of the dual Lie group $G^\ast$}

Now we need a faithful representation ${ \rho: g^\ast \to gl(N)}$ of the dual Lie algebra $g^\ast$: 
\be
[ \rho (x^j), \rho(x^k) ] =f^{jk}_i \rho(x^i).
\ee
In some cases the adjoint representation is faithful, and this simplifies the programming of the algorithm, although in general that could not be the case.
We use coordinates $X_i$ on $G^\ast$, so that a generic element of (the connected component of) the dual Lie group $G^\ast$ is obtained in the usual way:
\be
G^\ast (X_1,\dots, X_d)=G^\ast (\vec{X})= \prod_{i=1}^d \exp\left(X_i \rho(x^i) \right) .
\ee
In particular, we will use the following faithful representation of the book Lie algebra:
\be
\rho(j^+)=\left( 
\begin{array}{ccc}
0 & 0 & 1\\
0 & 0 & 0\\
0 & 0 & 0
\end{array}
\right)  \quad 
\rho(j^-)=\left( 
\begin{array}{ccc}
0 & 0 & 0\\
0 & 0 & 1\\
0 & 0 & 0
\end{array}
\right) \quad
\rho(j^3)=\left( 
\begin{array}{ccc}
-z & 0 & 0\\
0 & -z & 0\\
0 & 0 & 0
\end{array}
\right).
\label{repbook}
\ee
and we will denote the local group coordinates as $\{J_+,J_-,J_3 \}$, respectively.
The corresponding book Lie group element $G^\ast$ is given by:
\be
G^\ast=\exp \left(J_+ \rho(j^+) \right) \exp \left( J_- \rho(j^-) \right) \exp \left(J_3 \rho(j^3) \right)=
\left( 
\begin{array}{ccc}
 e^{-z J_3} & 0  & J_+\\
0 & e^{-z J_3} & J_-\\
0 & 0 & 1
\end{array}
\right).
\label{element}
\ee
Hereafter, the entries of the matrix $G^\ast$ will be denoted as $G^\ast_{ij}\ (i,j=1,\dots,N)$.

\subsection{The coproduct map $\Delta_{G^\ast}$}

The coproduct map for the coordinate functions on $G^\ast$ is just the pullback of the group law for the coordinate functions $X_i$ (see~\cite{BBMpullback}), but written in the (dual) algebraic form
\be
\Delta_{G^\ast}: C^\infty(G^\ast)\rightarrow C^\infty(G^\ast)\otimes  C^\infty(G^\ast).
\ee
In this language, the coassociativity constraint for $\Delta_{G^\ast}$
\be
(\Delta_{G^\ast}\otimes id)\circ \Delta_{G^\ast} = (id\otimes \Delta_{G^\ast})\circ \Delta_{G^\ast}
\ee
is nothing but the associativity of the group multiplication, provided that the entries of the first copy of the group element live on $C^\infty(G^\ast)\otimes 1$ and the second one lives on $1\otimes C^\infty(G^\ast)$. In this way, we can assume that the coproduct will be of the form
\begin{equation}
\Delta_{G^\ast}(X_i)=\sum_k g_{ik}(\vec{X}) \otimes h_{ik}(\vec{X}), \qquad i=1, \dots, d \label{copfact}
\end{equation}
where $\vec X=(X_1,\dots, X_d)$ and $g_{ik}$ and $h_{ik}$ will be certain smooth functions on $G^\ast$ that we have to find (the range of the index $k$ will depend on the number of different functions needed).
In particular, the explicit form of $\Delta_{G^\ast}$ will be obtained by solving the functional equations coming from imposing that the coproduct $\Delta_{G^\ast}(G^\ast_{ij})$ for the matrix entries of $G^\ast$ has to be given by the `tensorized' group multiplication in ${G^\ast}$, namely
\be
\Delta_{G^\ast}(G^\ast_{ij}(\vec{X}) ) =\sum_{l=1}^N G^\ast_{il}(\vec{X}) \otimes G^\ast_{lj}(\vec{X}) 
\label{feqs}
\ee
where $G_{ij}(\vec{X})$ is the corresponding entry of the group element $G^\ast$ in the chosen $N$-dimensional representation.

In the present book group example, the nine functional equations (\ref{feqs}) coming from the multiplication of two group elements (\ref{element}) can be written in matrix form as
\bea
&&
\left( 
\begin{array}{ccc}
 \Delta_{G^\ast}(e^{-z J_3}) &  \Delta_{G^\ast}(0)  & \Delta_{G^\ast}(J_+)\\
\Delta_{G^\ast}(0)  & \Delta_{G^\ast}(e^{-z J_3}) & \Delta_{G^\ast}(J_-)\\
\Delta_{G^\ast}(0)  & \Delta_{G^\ast}(0)  & \Delta_{G^\ast}(1)
\end{array}
\right)
=\cr
&&\qquad\qquad
=\left( 
\begin{array}{ccc}
 e^{-z  (J_3 \otimes 1)} & 0  & J_+\otimes 1\\
0 & e^{-z (J_3 \otimes 1)} & J_-\otimes 1\\
0 & 0 & 1\otimes 1
\end{array}
\right)
\cdot\left( 
\begin{array}{ccc}
 e^{-z (1\otimes J_3)} & 0  & 1\otimes J_+\\
0 & e^{-z (1\otimes J_3)} & 1\otimes J_-\\
0 & 0 & 1\otimes 1
\end{array}
\right)=\cr
&&\qquad\qquad
=
\left( 
\begin{array}{ccc}
 e^{-z J_3}\otimes e^{-z J_3}& 0  &  e^{-z J_3}\otimes J_+ + J_+ \otimes 1\\
0 &  e^{-z J_3}\otimes e^{-z J_3} & e^{-z J_3}\otimes J_- + J_- \otimes 1\\
0 & 0 & 1\otimes 1
\end{array}
\right)
\label{matrixp}
\eea
from which we immediately obtain the solution for the coproduct for the group coordinates $J_\pm$, namely 
\be
\Delta_{G^\ast} \left( J_\pm \right) = e^{-z J_3} \otimes J_\pm + J_\pm \otimes 1
\label{cojpm}
\ee
and since $\Delta_{G^\ast}(e^{-z J_3})=e^{-z J_3}\otimes e^{-z J_3}$ we finally have
\be 
\Delta_{G^\ast}(J_3)=1 \otimes J_3 + J_3 \otimes 1
\label{coj3}
\ee
as well as $ \Delta_{G^\ast}(1)=1\otimes 1$.

Evidently, in this naive example this solution can be straightforwardly obtained by hand,  but in a general case we know that the solution for the functional equations (\ref{feqs}) does always exist (it is just the group law for the group under consideration). In fact, for all the examples we will provide in this paper, as well as for many others we have considered, this problem can be fully solved by computer methods. Nevertheless, it is true that in some cases these functional equations could be too complicated to be solved by a symbolic manipulation program. 

At this point it is worth stressing that, essentially, the exponentiation of the dual Lie algebra $g^\ast$ giving rise to the dual Lie group $G^\ast$ can be understood as the canonical way to `exponentiate' the first order deformation of the coproduct given by $\delta$ to the one corresponding to the full quantum deformation. Therefore, any coproduct can be thought of as the multiplication law on the appropriate dual Lie group (in this case, the book group). Note that the choice of different local coordinate functions on the group gives different expressions for the same abstract comultiplication law on $G^\ast$.

\subsection{The PL bracket $\Lambda$ on $G^\ast$}

We recall that the unique PL group structure on $G^\ast$ which is in one to one correspondence with $(g^\ast,\delta^\ast)$ via the Drinfel'd theorem~\cite{Dri} has to fulfill two conditions: firstly, the group (co)multiplication has to be a Poisson map with respect to $\Lambda$:
\be
\pois{\Delta_{G^\ast} (a)}{\Delta_{G^\ast} (b)}_\Lambda=\Delta_{G^\ast}(\pois{a}{b}_\Lambda) \qquad
\forall\,a,b\,\in C^\infty(G^\ast)
\label{homo}
\ee
and, secondly, the linearization of $\Lambda$ should coincide with the Lie algebra defined by the structure tensor $c_{ij}^k$ that defines $\delta^\ast$. 

At this point we have two possibilities in order to compute $\Lambda$:

\noindent $\bullet$ a) If the dual Lie bialgebra $(g^\ast,\delta^\ast)$ is a {\em coboundary} one (i.e. it comes from a classical $r$-matrix on $g^\ast\otimes g^\ast$) then the Poisson structure $\Lambda$ is given by the Sklyanin bracket on $G^\ast$
\be
\{a,b\}_\Lambda=r^{ij} \left( X^L_i (a) \,X^L_j (b) -  X^R_i (a) X^R_j (b) \right) \qquad a,b \in C^\infty(G^\ast),
\ee
where $X^L_i, \ X^R_i, (i=1,\dots,d)$ are, respectively, left and right invariant vector fields on $G^\ast$ and $r^{ij}$ are the components of the $r$-matrix in the chosen basis $\{X_1,\dots, X_d\}$.

\noindent $\bullet$  b)  If the dual Lie bialgebra $(g^\ast,\delta^\ast)$ is a {\em non-coboundary} one, one could try to find 
the PL structure on $G^\ast$ analytically, but this is a quite cumbersome procedure (see~\cite{euclideo,galileo,anna,brihaye,checos}). Here we will propose to use the computer assisted method that has been recently used in~\cite{BBM3d} to provide all real 3D PL groups.  As we will see in the sequel, this algorithm will work in a very efficient way, and we will easily obtain the explicit PL structures we need even in the case of high dimensional Lie groups.

This second method will be based on the assumption that the PL brackets for the local coordinates on $G^\ast$ are of the form
\be
\pois{X_i}{X_j}_\Lambda=\sum_{k,l} \beta_{ijkl} F_k F_l 
\label{ansatz}
\ee
with arbitrary unknown coefficients $\beta_{ijkl}$ and with $F_i$ being any of the functions
\begin{equation}
{\vec{F}}= \left\{G_{ij}^\ast(\vec{X}), g_{im}(\vec{X}), h_{im} (\vec{X}) \right\}
\label{ansatz2}
\end{equation}
which includes only the $G_{ij}^\ast(\vec{X})$ functions appearing as matrix entries of the group element $G^\ast$, as well as all the $g_{im}(\vec{X})$ and $h_{im} (\vec{X})$ functions appearing in the coproducts for the local coordinates (\ref{copfact}). This means that we are looking for Poisson brackets for the local coordinates that are homogeneous quadratic in terms of functions included within the set $\vec{F}$. 

It is important to stress that if the dual Lie bialgebra is a coboundary one, the Sklyanin bracket guarantees that the PL structure will be homogeneous quadratic in the matrix entries $G_{ij}^\ast(\vec{X})$. Therefore our Ansatz is a slight generalization of this fact, that turns out to be suitable for dealing also with non-coboundary PL structures~\cite{BBM3d}. Note that the fact that the Ansatz (\ref{ansatz}) works is crucial for the algorithmic construction of the PL group $G^*$. Although this is not guaranteed in general, we recall that this Ansatz works for constructing all the PL structures on any 3D real Lie group, as shown in~\cite{BBM3d}, and indeed it will work for all the cases here presented.

Therefore, let us construct the antisymmetric $d \times d$ matrix with entries $Q_{ij}$ given by
\be
\pois{X_i}{X_j}_\Lambda=Q_{ij}=\sum_{k,l} \beta_{ijkl} F_k F_l .
\label{qij}
\ee
Then, the homomorphism condition (\ref{homo}) for the coproduct $\Delta_{G^\ast}$ is translated into
\bea
&&\!\!\!\!\!\!\!\!\!\!\!\!\!\!\!\!\!\!\!\!\!\!
\{ \Delta_{G^\ast}(X_i), \Delta_{G^\ast}(X_j) \}_\Lambda= \\
&&\!\!\!\!\!\!\!\!\!\!\!\!\!\!\!\!
\sum_{k,l=1}^{d} \left( Q_{kl} \otimes 1 \right) \frac{\partial \Delta_{G^\ast}(X_i)}{\partial (X_k \otimes 1)} \cdot \frac{\partial \Delta_{G^\ast}(X_j)}{\partial (X_l \otimes 1)}+
\left( 1 \otimes Q_{kl} \right) \frac{\partial \Delta_{G^\ast}(X_i)}{\partial (1 \otimes X_k)} \cdot \frac{\partial \Delta_{G^\ast}(X_j)}{\partial (1 \otimes X_l)}=\Delta_{G^\ast}(Q_{ij})
\label{cophom}
\eea
which gives rise to a set of  linear equations for the coefficients $\beta_{ijkl}$. After solving this system of equations, we have to impose the linearization condition
\be
\left. \sum_{k=1}^d \frac{\partial Q_{ij}}{\partial X_k}  \right|_{X_1=X_2=\dots,X_d=0} X_k= c_{ij}^k X_k 
\ee
on the remaining coefficients $\beta_{ijkl}$. If the Ansatz (\ref{ansatz}) concerning the functions works, it turns out that the solution for the coefficients $\beta_{ijkl}$ is unique, and $Q$ gives the Poisson 
tensor $\Lambda$ for the PL group associated with ${(g^\ast,\delta^\ast)}$. Indeed, in this way  the Jacobi identity for $Q$ will be automatically satisfied. 
In particular, the linearization constrain would imply that in the vanishing limit of the deformation parameter(s) we recover the Poisson analogue of the initial Lie algebra $g$:
\be
\lim_{z \to 0}\pois{X_i}{X_j}_\Lambda=\lim_{z \to 0} Q_{ij}=c_{ij}^k X_k.
\ee
We recall that this algorithm has been sucessfully used in~\cite{BBM3d} to construct the complete set of three-dimensional PL group structures for all three-dimensional real Lie groups.

In the particular case we are solving now, the Ansatz (\ref{ansatz}) will imply that the PL bracket $\Lambda$ is at most quadratic in the following functions  
\be
{\vec{F}}:= \{1, J_3, J_+, J_-,  e^{-z J_3}\}.
\ee
Therefore, we obtain a matrix $Q$ depending on $d(d-1)/2 \times s(s+1)/2=45$ free parameters $\beta_{ijkl}$, where $d$ is the dimension
of the algebra ($d=3$ in this case) and $s$ is the number of elements in the set of functions we have chosen in order
to construct our $Q_{ij}$ matrix ($s=5$ in this case). Thus, we have to impose that the bracket defined by  (recall that we have chosen $X_1=J_+, X_2=J_-, X_3=J_3$)
\be
\{ X_i, X_j \}_\Lambda= Q_{ij}
\ee
satisfies the coproduct homomorphism property (\ref{cophom}), and in this way we obtain a set of linear equations for our $45$ free parameters. After solving them, our matrix elements $Q_{ij}$ become 
\begin{eqnarray}
Q_{12}&=&\{ J_+, J_- \}\ = \beta_{1, 2, 5, 5} \left[ \exp(-2 z J_3)-1 \right]+ \frac{z}{2} \beta_{2, 3, 1, 3} J_+^2 - z \beta_{2, 3, 1, 5} J_+ \nonumber\\
&&\qquad\qquad\qquad   +z \beta_{2, 3, 1, 4} J_+ J_- 
- \frac{z}{2} \beta_{1, 3, 1, 4} J_-^2+z \beta_{1, 3, 1, 5} J_- .
\nonumber \\
Q_{13}&=& \{ J_+, J_3 \} \ = \beta_{1, 3, 1, 5} \left[ \exp(-z J_3)-1 \right] +\beta_{1, 3, 1, 4} J_- - \beta_{2, 3, 1, 4} J_+ \label{nolineal}\\
Q_{23}&=& \{ J_-, J_3 \} \ =\beta_{2, 3, 1, 5} \left[ \exp(-z J_3)-1 \right] +\beta_{2, 3, 1, 3} J_+ + \beta_{2, 3, 1, 4} J_-   \nonumber 
\end{eqnarray}
Therefore, we have a multiparametric family of pre-Poisson brackets (the Jacobi identity has not been imposed yet, and actually it is not satisfied at this stage) that are compatible with the group (co)multiplication  on $G^\ast$. Nevertheless, Drinfel'd  theorem says that only one of them corresponds to the Lie bialgebra $(g^\ast,\delta^\ast)$. In order to identify such unique solution, 
we write the linearization $\Lambda_0$ (with respect to the group coordinates) of the bracket (\ref{nolineal}), which reads:
\begin{eqnarray}
\{ J_3, J_+ \}_0 &=&  z \beta_{1, 3, 1, 5} J_3+\beta_{2, 3, 1, 4} J_+ -\beta_{1, 3, 1, 4} J_- \nonumber \\
\{ J_3, J_- \}_0 &=& z \beta_{2, 3, 1, 5} J_3- \beta_{2, 3, 1, 3} J_+ -\beta_{2, 3, 1, 4} J_- \\
\{ J_+, J_- \}_0 &=& -2 z \beta_{1, 2, 5, 5} J_3+z \beta_{1, 3, 1, 5} J_+ - z \beta_{2, 3, 1, 5} J_-.
\nonumber
\end{eqnarray}
Thus, the unique PL bracket associated to $(g^\ast,\delta^\ast)$ would be the one whose linearized part reproduces the $c_{ij}^k$ tensor, {\em i.e.}, the original $sl(2,\mathbb{R})$ Lie algebra structure 
\begin{equation}
[j_3,j_\pm]=\pm 2 j_\pm, \quad [j_+,j_-]=j_3,
\end{equation}
under the identification $j_3\equiv J_3,\  j_+\equiv J_+,\  j_-\equiv J_-$ (that comes from the way in which we have defined the dual Lie bialgebra).
Obviously, this means that we have to impose the relations
\be
\beta_{1, 3, 1, 4}=\beta_{1, 3, 1, 5}=\beta_{2, 3, 1, 3}=\beta_{2, 3, 1, 5}=0
\ee
\be
\beta_{2, 3, 1, 4}=2 \qquad
\beta_{1, 2, 5, 5}=-\frac{1}{2z}
\ee
and the unique solution for the PL bracket $\Lambda$ turns out to be:
\be
\{ J_3, J_\pm \}_\Lambda=\pm 2 J_\pm, \qquad \{ J_+, J_- \}_\Lambda=\frac{ 1-\exp(-2 z J_3)}{2z}+2 z\, J_+ J_- 
\ee
The one-to-one correspondence between PL groups and Lie bialgebra structures implies that the Jacobi identity for this bracket is automatically fulfilled and,  together with the coproduct (\ref{cojpm})-(\ref{coj3}), it defines the Poisson-Hopf algebra which is in one to one correspondence with the dual Lie bialgebra $(g^\ast,\delta^\ast)$. We remark that, througout the paper, we will not present counit and antipode maps, that can be straightforwardly deduced from the coproduct map.

\subsection{Quantization}

The Hopf algebra quantization of this dual PL group will be just the quantum algebra we are looking for.  Note that the above Poisson brackets present ordering problems (coming from the product $J_+J_-$) that would make evident when we want to transform the group coordinates into non-commutative generators. Nevertheless, if we perform the following change of local coordinates on the group $G^\ast$
\be
J_\pm'=e^{\frac{z}{2} J_3} J_\pm
\label{newcoord}
\ee
the PL group structure reads
\begin{eqnarray}
\Delta_{G^\ast}(J_3)&=&J_3 \otimes 1 + 1 \otimes J_3 \label{coin1}\\
\Delta_{G^\ast}(J_\pm')&=& J_\pm' \otimes e^{\frac{z}{2} J_3} +  e^{-\frac{z}{2} J_3} \otimes J_\pm'
\label{coin2}
\end{eqnarray}  
\be
\{ J_3, J_\pm' \}=\pm 2 J_\pm', \qquad \{ J_+', J_-' \}=\frac{\sinh\left( z J_3 \right)}{z} 
\ee
and the ordering ambiguities disappear. Note that the linearization of this bracket (that coincides with the $z\rightarrow 0$ limit) gives
\be
\{ J_3, J_\pm' \}=\pm 2 J_\pm', \qquad \{ J_+', J_-' \}= J_3  ,
\ee
which is again the Poisson-$sl(2,\mathbb{R})$ algebra given by the $c_{ij}^k$ tensor.

Now, the quantization prescription would consist essentialy in substituting the Poisson brackets by commutators
$
\{\cdot,\cdot\}\rightarrow [\cdot,\cdot]
$ and to map the dual group coordinates into non-commutative algebra generators $X_i\rightarrow \hat X_i$ both in the coproduct and in the commutation rules.
In this way we obtain the well-known Drinfel'd-Jimbo quantum deformation of the $sl(2,\mathbb{R})$ algebra
\begin{eqnarray}
\Delta_z(\hat J_3)&=&\hat J_3 \otimes 1 + 1 \otimes \hat J_3 \\
\Delta_z(\hat J_\pm')&=& \hat J_\pm' \otimes e^{\frac{z}{2} \hat J_3} +  e^{-\frac{z}{2} \hat J_3} \otimes \hat J_\pm'
\end{eqnarray}  
\be
[\hat J_3, \hat J_\pm' ]=\pm 2 \hat J_\pm', \qquad [\hat J_+', \hat J_-' ]=\frac{ \sinh( z \hat J_3 ) }{z} 
\ee
for which the homomorphism condition on the coproduct 
\be
[\Delta_z (\hat X),\Delta_z (\hat Y)]=\Delta_z\left([\hat X,\hat Y]\right) 
\label{homoquant}
\ee
can be straightforwardly checked.

We could summarize this construction by saying that the Poisson analogue of this $q$-defor\-mation of $sl(2,\mathbb{R})$ is the specific PL structure $\Lambda$ on the book group, and the $q$-deformed coproduct is just the  pullback of the book group product law in a specific set $\{J_3,J_\pm'\}$ of local coordinates. Note that  the  construction of Poisson $sl_q(2)$ as a PL group structure on $SB(2,C)$ was already given in~\cite{Marmo}.

\noindent {\bf Remark 1.} We stress that in this approach the limit $z \to 0$ corresponds to the abelian group law for the group $G^\ast$, which is tantamount to a `non-deformed' (primitive) coproduct for the group coordinates. In other words, if we consider any Lie algebra $g$ endowed with the trivial Lie bialgebra structure $\delta=0$, the dual Lie bialgebra $g^\ast$ will be the abelian Lie algebra. Therefore, the group law for the abelian group $G^\ast$ will induce the primitive coproduct $\Delta_{G^\ast} ( X)=1\otimes X + X\otimes 1$ for all the abelian group coordinates, and the quantization of the PL structure that corresponds to $\delta^\ast$ will be just the (non-deformed) universal enveloping algebra $U(g)$ with the primitive Hopf algebra structure given by $\Delta_0 (X)=1\otimes X + X\otimes 1$.

\noindent {\bf Remark 2.} It is important to realize that the coordinates (\ref{newcoord}) in which the ordering ambiguities disappear are such that their coproduct map (\ref{coin1})-(\ref{coin2}) is invariant under the composition of the flip operator action $\sigma(X\otimes Y)=Y\otimes X$ and the reversal $z\to -z$ of the quantum deformation parameter. This seems to be a general property that will hold in the rest of the examples here presented, and this fact can be used as a guiding principle that can be easily implemented in order to find the appropriate `ambiguity-free' coordinates. Note that, after quantization, since the quantum universal $R$ matrix for a given $U_z(g)$ is the invertible operator $R\in U_z(g)\otimes U_z(g)$ such that 
\be
R\circ \Delta_z \circ R^{-1}= \sigma\circ \Delta_{z},
\label{actionr}
\ee
this means that (in the case that such a quantum $R$ matrix operator does exist) the coordinates  (\ref{newcoord}) are the ones whose quanization give rise to a quantum algebra  for which the action of $R$ on $\Delta_z$ given by (\ref{actionr}) coincides with the change $z\to -z$ in the expressions for $\Delta_z$.


\section{A quantum $so(2,2)$ algebra}

As a much more complicated example from the computational viewpoint, let us consider the Anti de Sitter Lie algebra $so(2,2)$ with the following generators $\{n_-, n_+, n_3, j_-,j_+,j_3\}$ and the 
commutation rules:
\begin{eqnarray}
&&\left[ j_3,j_{\pm} \right]=\left[ n_3,n_{\pm} \right]=\pm 2 j_{\pm}\nonumber \\
&&\left[ j_3, n_{\pm} \right] = \left[ n_3, j_{\pm} \right]=\pm 2 n_{\pm}\nonumber \\
&&\left[ j_+, j_{-} \right] = \left[ n_+, n_{-} \right]=j_3 \label{so22}\\
&&\left[ j_{\pm}, n_{\mp} \right]= \pm n_3 .
\nonumber
\end{eqnarray}
Now, let us show that the (one-parameter) Drinfel'd-Jimbo deformation of $so(2,2)$ can be explicitly obtained through the method described in the previous Section. The initial Lie bialgebra $(g,\delta)$ that characterizes this deformation is (see~\cite{beyond})
\be
r=\frac{z }{2} \left( j_+ \wedge n_{-} + n_{+} \wedge j_{-} \right),
\ee
and the corresponding cocommutator $\delta$ is given by:
\begin{eqnarray}
\delta(j_3)&=&0 \cr
\delta(j_+)&=&-\frac{z }{2} \left(j_3 \wedge n_+ +n_3 \wedge j_+ \right)\nonumber \\
\delta(j_-)&=&-\frac{z }{2} \left( j_3 \wedge n_- + n_3 \wedge j_- \right) \nonumber \\
\delta(n_3)&=&0\\
\delta(n_+)&=&-\frac{z }{2} \left( j_3 \wedge j_+ + n_3 \wedge n_+ \right)\nonumber \\
\delta(n_-)&=&-\frac{z }{2} \left( j_3 \wedge j_- + n_3 \wedge n_- \right).
\nonumber
\end{eqnarray}  

Let $\{n^-, n^+, n^3, j^-,j^+,j^3\}$ be the basis of the dual Lie bialgebra $(g^\ast,\delta^\ast)$. The Lie algebra relations for $g^\ast$ are given by `dualizing' the initial cocommutator $\delta$, and read:
\begin{eqnarray}
&&\left[ j^3,j^{\pm} \right]=\left[ n^3,n^{\pm} \right]=-\frac{z }{2} n^{\pm}\nonumber \\
&&\left[ j^3, n^{\pm} \right] = \left[ n^3, j^{\pm} \right]=-\frac{z }{2} j^{\pm}\nonumber \\
&&\left[ j^+, j^{-} \right] = \left[ n^+, n^{-} \right]=0\\
&&\left[ j^{\pm}, n^{\mp} \right]=0 .
\nonumber
\end{eqnarray}
This is a six-dimensional solvable Lie algebra. It is straightforward to check that the dual Lie bialgebra $(g^\ast,\delta^\ast)$ (with $\delta^\ast$ obtained through the `dualization' of the $so(2,2)$ commutation rules) is a non-coboundary one, since a classical $r$-matrix for it does not exist. Therefore, we will be forced to use the second of the two alternatives given in section 2.5 for the computation of the dual PL group.

Now, a generic element of $G^\ast$ can be obtained through the exponentiation of the adjoint representation $\rho$ of this $g^\ast$, where the local coordinates will be given by $\{N_-, N_+, N_3, J_-,J_+,J_3\}$:
\be
G^\ast=\exp \left(N_{-} \rho(n^{-}) \right) \exp \left( N_+ \rho(n^+) \right) \exp \left(N_3 \rho(n^3) \right)
\exp \left(J_- \rho(j^-) \right) \exp \left( J_+ \rho(j^+) \right) \exp \left(J_3 \rho(j^3) \right),
\ee
namely
\footnotesize{
\be
\!\!\!\!\!\!\!\!\!\!\!\!\!\!\!\!
G^\ast=\left( 
\begin{array}{ccccccc}
1 & 0 & 0 & 0 & 0 & 0\\
\frac{z }{2} N_+  & e^{-\frac{z }{2} N_3} \cosh \left(\frac{z }{2} J_3 \right) & 0 &\frac{z }{2} J_+ e^{-\frac{z }{2} N_3} & -e^{-\frac{z }{2} N_3} \sinh \left(\frac{z }{2} J_3 \right) & 0\\ 
\frac{z }{2} N_- & 0 &e^{-\frac{z }{2} N_3} \cosh \left(\frac{z }{2} J_3 \right) & \frac{z }{2} J_- e^{-\frac{z }{2} N_3} & 0 & -e^{-\frac{z }{2} N_3} \sinh \left(\frac{z }{2} J_3 \right) \\ 
0 & 0 & 0 & 1 & 0 & 0\\ 
\frac{z }{2} J_+ e^{-\frac{z }{2} N_3} &-e^{-\frac{z }{2} N_3} \sinh\left( \frac{z }{2} J_3 \right) & 0 & \frac{z }{2} N_+& e^{-\frac{z }{2} N_3} \cosh \left(\frac{z }{2} J_3 \right)  & 0\\ 
\frac{z }{2} J_- e^{-\frac{z }{2} N_3} & 0 &-e^{-\frac{z }{2} N_3} \sinh\left( \frac{z }{2} J_3 \right)& \frac{z }{2} N_- & 0 & e^{-\frac{z }{2} N_3} \cosh \left(\frac{z }{2} J_3 \right) 
\end{array}
\right).
\ee}\normalsize

By solving the functional equations arising from the multiplication of two group elements of this type, we get the following coproduct map for the local coordinate functions:
\begin{eqnarray}
\Delta_{G^\ast}\left( J_3\right) &=& J_3 \otimes 1 +1 \otimes J_3 \nonumber \\
\Delta_{G^\ast} \left( J_{\pm} \right) &=& \cosh \left(\frac{z }{2} J_3 \right) \otimes J_{\pm}+ J_{\pm} \otimes e^{\frac{z }{2} N_3} - \sinh \left(\frac{z }{2} J_3 \right) \otimes N_{\pm} e^{\frac{z }{2} N_3} \nonumber \\
\Delta_{G^\ast}\left( N_3\right)&=&N_3 \otimes 1 + 1 \otimes N_3 \\
\Delta_{G^\ast} \left( N_{\pm} \right) &=& N_{\pm} \otimes 1 + e^{-\frac{z }{2} N_3} \cosh \left(\frac{z }{2} J_3 \right) \otimes N_{\pm} - e^{-\frac{z }{2} N_3} \sinh \left(\frac{z }{2} J_3 \right) \otimes J_{\pm} e^{-\frac{z }{2} N_3} .
\nonumber
\end{eqnarray}

Now, we want to construct the Poisson tensor compatible with $\Delta_{G^\ast}$ that gives the unique PL structure on $G^\ast$ whose associated Lie bialgebra is $(g^\ast,\delta^\ast)$. As described in the previous Section, 
our Ansatz will be that such Poisson tensor is quadratic in the functions of the group coordinates that appear in the group entries and in the coproducts of the coordinate functions.
In this case the set $\vec F$ contains 16 different functions, namely
\begin{eqnarray}
&&\!\!\!\!\!\!\!\!\!\!
\vec{F}:=\left\{ 1, J_3,J_+,J_-,N_3,N_+,N_-,\cosh \left(\frac{z }{2} J_3 \right),\sinh \left(\frac{z }{2} J_3 \right), \right.\nonumber \\
&&\!\!\!\!\!\!\!\!\!\!\!\!\!\!\!\!\!\!\!\! \left. \qquad e^{\frac{z }{2} N_3}, N_+ e^{\frac{z }{2} N_3}, N_- e^{\frac{z }{2} N_3},
e^{-\frac{z }{2} N_3}  \cosh \left(\frac{z }{2} J_3 \right), e^{-\frac{z }{2} N_3}  \sinh \left(\frac{z }{2} J_3 \right), J_+ e^{-\frac{z }{2} N_3}, J_- e^{-\frac{z }{2} N_3} \right\}.
\end{eqnarray}
Consequently, we construct the most general antisymmetric matrix $Q_{ij}$ being quadratic
in these variables and we require $\Delta_{G^\ast}$ to be a Poisson morphism with respect to the (tentative) Poisson bracket
defined by $Q$ (\ref{qij}). 
This requirement produces a system
of linear equations inluding $2040$ $\beta_{ijkl}$ coefficients, that can be solved by using a symbolic manipulation program.
The result gives us the most general quadratic matrix compatible with the coproduct $\Delta_{G^\ast}$ (this matrix still do not 
define a Poisson tensor because we have not imposed it to fulfill the Jacobi identity yet). Afterwards, if we impose that the linearization
of this tentative Poisson tensor corresponds to the structure tensor $c_{ij}^k$ that defines the initial $so(2,2)$ Lie algebra (\ref{so22}), we get the following unique solution (we give the nonvanishing Poisson brackets only): 
\begin{eqnarray}
\left\{ J_3,J_{\pm} \right\}_\Lambda&=&\pm 2 J_{\pm} \nonumber \\
\left\{ J_3,N_{\pm} \right\}_\Lambda&=&\pm 2 N_{\pm} \nonumber \\
\left\{ J_+, J_{-} \right\}_\Lambda&=&\frac{\sinh \left( z   J_3 \right)}{z } \nonumber \\
\left\{ J_+, N_{3} \right\}_\Lambda &=& -2 N_+ e^{\frac{z }{2} N_3} \nonumber \\
\left\{ J_+, N_+ \right\}_\Lambda &=& z  J_+^2 e^{-\frac{z }{2} N_3} \nonumber \\
\left\{ J_+, N_{-} \right\}_\Lambda&=& \frac{1}{z } \left[e^{\frac{z }{2} N_3} \left( 1 + z ^2 N_+ N_{-} \right) - e^{-\frac{z }{2} N_3} \cosh \left( z   J_3 \right) \right] \\
\left\{ J_-, N_3 \right\}_\Lambda&=& 2 N_- e^{\frac{z }{2} N_3} \nonumber \\
\left\{ J_-, N_{+} \right\}_\Lambda&=& -\frac{1}{z } \left[e^{\frac{z }{2} N_3} \left( 1 + z ^2 N_+ N_{-} \right) - e^{-\frac{z }{2} N_3} \cosh \left( z   J_3 \right) \right]  \nonumber \\
\left\{ J_-, N_- \right\}_\Lambda &=& -z  J_-^2 e^{-\frac{z }{2} N_3} \nonumber \\
\left\{ N_3,N_{\pm} \right\}_\Lambda&=&\pm 2 J_{\pm} e^{-\frac{z }{2} N_3} \nonumber \\
\left\{ N_+, N_{-} \right\}_\Lambda&=& \frac{1}{z } \left[z ^2 e^{-\frac{z }{2} N_3} \left(J_+ N_{-} + N_+ J_- \right) + e^{-\frac{z }{2} N_3} \sinh \left( z   J_3 \right) \right]
\nonumber
\end{eqnarray}
that can be easily proven to fulfill the Jacobi identity and gives us the Poisson analogue of the $so(2,2)$ quantum deformation we were looking for.

Therefore, the final step would be to quantize this Poisson-Hopf algebra, but we observe that both in the coproduct and -much more strongly- in the Poisson brackets, we have hard ordering problems. Nevertheless, all of them can be circumvented by defining the following change of (still commutative) coordinates 
\begin{eqnarray}
J_\pm'&=&\exp\left( -\frac{z }{4} N_3 \right) \left[ \cosh \left( \frac{z }{4} J_3 \right)J_\pm+ \sinh \left( \frac{z }{4} J_3 \right) \exp\left( \frac{z }{2} N_3 \right) N_\pm \right], \nonumber \\
N_\pm'&=&\exp\left( -\frac{z }{4} N_3 \right) \left[ \cosh \left( \frac{z }{4} J_3 \right) \exp\left( \frac{z }{2} N_3 \right) N_\pm+ \sinh \left( \frac{z }{4} J_3 \right)J_\pm \right].
\end{eqnarray}
In these new coordinates, the new non-vanishing Poisson brackets read
\begin{eqnarray}
\left\{ J_3,J_{\pm}' \right\}_\Lambda&=&\left\{ N_3,N_{\pm}' \right\}=\pm 2 J_{\pm}' \nonumber \\
\left\{ J_3,N_{\pm}' \right\}_\Lambda&=&\left\{ N_3,J_{\pm}' \right\}=\pm 2 N_{\pm}' \nonumber \\
\left\{ J_+', J_{-}' \right\}_\Lambda&=& \left\{ N_+', N_{-}' \right\}=\frac{2}{z } \sinh \left( \frac{z }{2} J_3 \right) \cosh \left( \frac{z }{2} N_3 \right)   \\
\left\{ N_{3}, J_{\pm}' \right\}_\Lambda &=& \pm 2 N_{\pm}'  \nonumber \\
\left\{ J_\pm', N_{\mp}' \right\}_\Lambda&=& \frac{2}{z } \sinh \left( \frac{z }{2} N_3 \right) \cosh \left( \frac{z }{2} J_3 \right) 
\nonumber
\end{eqnarray}
and the form of the new coproduct is:
\begin{eqnarray}
\Delta_{G^\ast}\left(J_3\right) &=& J_3 \otimes 1 +1 \otimes J_3 \nonumber \\
\Delta_{G^\ast} \left(J_{\pm}' \right) &=&J_{\pm}' \otimes e^{\frac{z }{4} N_3} \cosh \left( \frac{z }{4} J_3 \right) + e^{-\frac{z }{4} N_3} \cosh \left( \frac{z }{4} J_3 \right) \otimes J_{\pm}' + \nonumber \\
&& \qquad\qquad
N_{\pm}' \otimes e^{\frac{z }{4} N_3} \sinh \left( \frac{z }{4} J_3 \right) - e^{-\frac{z }{4} N_3} \sinh \left( \frac{z }{4} J_3 \right) \otimes N_{\pm}'  \nonumber \\
\Delta_{G^\ast}\left(N_3\right)&=&N_3 \otimes 1 + 1 \otimes N_3 \\
\Delta_{G^\ast} \left(N_{\pm}' \right) &=& N_{\pm}' \otimes e^{\frac{z }{4} N_3} \cosh \left( \frac{z }{4} J_3 \right) + e^{-\frac{z }{4} N_3} \cosh \left( \frac{z }{4} J_3 \right) \otimes N_{\pm}' + \nonumber \\
&& \qquad\qquad
J_{\pm}' \otimes e^{\frac{z }{4} N_3} \sinh \left( \frac{z }{4} J_3 \right) - e^{-\frac{z }{4} N_3} \sinh \left( \frac{z }{4} J_3 \right) \otimes J_{\pm}' .
\nonumber
\end{eqnarray}
Since $J_3$ and $N_3$ Poisson-commute, we would have no ordering ambiguities at all if we define the quantization by substituting Poisson brackets by commutators
$
\{\cdot,\cdot\}\rightarrow [\cdot,\cdot]
$ and by mapping the new coordinate functions to noncommutative generators both in the coproduct and in the commutation rules. Note that the new coordinate functions are such that their coproduct fulfills the symmetry property stated in Remark 2 of section 3.  We omit the explicit writing of these truly `quantum' relations, since they are exactly the ones obtained for the Drinfel'd-Jimbo quantization of $so(2,2)$ in~\cite{beyond}.


\section{The (3+1) $\kappa$-Poincar\'e algebra}

The next example we are going to consider is a 10-dimensional one, and consists in the Hopf algebra deformation leading to the well-known $\kappa$-Poincar\'e algebra. As we will see, the dual PL group construction will naturally give rise to a Poisson analogue of the $\kappa$-Poincar\'e algebra that is written in the bicrossproduct basis~\cite{Majidbicross}.

The set of generators for the (3+1) Poincar\'e algebra is taken as $\{p_0,p_1,p_2,p_3,k_1,k_2,k_3,j_1,j_2,j_3 \}$ (translations $p_i$, boosts $k_i$, rotations $j_i$) and their commutation 
relations read:
\be
\begin{array}{lll}
\left[ j_i,j_j \right]=\epsilon_{ijk} j_k & \left[ j_i,p_{j} \right]=\epsilon_{ijk} p_k & \left[ j_i,k_{j} \right]=\epsilon_{ijk} k_k\\[2pt]
\left[ p_i, p_j \right]=0 & \left[ p_i, k_j \right]=-\delta_{ij} p_0 & \left[ k_i, k_j \right]=-\epsilon_{ijk} j_k\\[2pt]
\left[ p_0, p_i \right]=0 &  \left[ p_0, k_{i} \right]=-p_i & \left[ p_0, j_i \right]= 0 .
\end{array}
\ee
The so-called $\kappa$-deformation is generated by the following classical $r-$matrix:
\be
r=z \left( k_1 \wedge p_1 + k_2 \wedge p_2+ k_3 \wedge p_3 \right)
\ee
from which the cocommutator $\delta$ is given by:
\begin{eqnarray}
\delta(p_0)&=&0 \nonumber \\
\delta(p_1)&=&z \ p_1 \wedge p_0    \nonumber \\
\delta(p_2)&=&z \ p_2 \wedge p_0   \nonumber \\
\delta(p_3)&=&z \ p_3 \wedge p_0   \nonumber \\
\delta(k_1)&=&z \left(k_1 \wedge p_0 + p_2 \wedge j_3 + j_2 \wedge p_3 \right)  \nonumber \\
\delta(k_2)&=&z \left( j_3 \wedge p_1 + p_3 \wedge j_1 + k_2 \wedge p_0 \right) \\
\delta(k_3)&=&z \left(p_1 \wedge j_2 + j_1 \wedge p_2  + k_3 \wedge p_0 \right)  \nonumber \\
\delta(j_1)&=&0 \nonumber \\
\delta(j_2)&=&0 \nonumber \\
\delta(j_3)&=&0.
\nonumber
\end{eqnarray}  

Let $\{p^0,p^1,p^2,p^3,k^1,k^2,k^3,j^1,j^2,j^3\}$ be the basis of the dual algebra $g^\ast$, whose Lie algebra structure is given by the structure constants appearing in the initial cocommutator $\delta$. Namely:
\be
\begin{array}{lll}
\left[ p^0,p^i \right]= -z\ p^i  &\left[ p^0,k^i \right]= -z\ k^i  &\left[ p^0,j^i \right]= 0 \\[2pt]
\left[ p^i,p^j \right]= 0  &\left[ p^i,k^j \right]= 0  &\left[ p^i,j^j \right]= z\ \epsilon_{ijl} k^l \\[2pt]
\left[ k^i,k^j \right]= 0  &\left[ k^i,j^j \right]= 0  &\left[ j^i,j^j \right]= 0.
\end{array}
\ee
This is a solvable Lie algebra and, again, a direct computation shows that its Lie bialgebra structure $(g^\ast,\delta^\ast)$, with $\delta^\ast$ coming from the structure tensor of the (3+1) Poincar\'e algebra, is a non-coboundary one. Therefore, The Poisson bracket on the dual group will be again obtained by using the same computational approach as in the previous examples.

Firstly, we have to find a faithful representation for $g^\ast$. The 10-dimensional adjoint one works, and  from it we construct the corresponding Lie group element by computing:
\begin{eqnarray}
\!\!\!\!\!\!\!\!\!\!\!\!\!\!\!\!
G^\ast&=&\exp \left(P_0 \rho(p^0) \right) \exp \left( P_1 \rho(p^1) \right) \exp \left(P_2 \rho(p^2) \right)
\exp \left(P_3 \rho(p^3) \right) \times \nonumber\\
&& \!\!\!\!\!\!\!\!\times \exp \left( K_1 \rho(k^1) \right) \exp \left(K_2 \rho(k^2) \right) \left(K_3 \rho(k^3) \right) \exp \left( J_1 \rho(j^1) \right) \exp \left(J_2 \rho(j^2) \right)
\exp \left(J_3 \rho(j^3) \right)
\end{eqnarray}
where $\{P_0,P_1,P_2,P_3,K_1,K_2,K_3,J_1,J_2,J_3 \}$ are the coordinate functions on $G^\ast$. The result for the group element is

{\footnotesize
\be
\!\!\!\!\!\!\!\!\!\!\!\!\!\!\!\!\!\!\!\!
G^\ast=
\left( 
\begin{array}{cccccccccc}
  1  & 0  &  0  & 0  &  0  & 0  & 0   & 0  &  0 & 0\\
  z e^{-zP_0} P_1  & e^{-zP_0}  & 0   & 0  &  0  & 0  &  0  & 0  & 0   &  0   \\
  z e^{-zP_0} P_2  & 0  & e^{-zP_0}   & 0  &  0  & 0  &  0  & 0  & 0   &  0 \\
  z e^{-zP_0} P_3   & 0  &  0  & e^{-zP_0}  & 0   & 0   & 0   & 0  & 0   & 0 \\
  z e^{-zP_0} K_1   & 0  & -z e^{-z P_0} J_3 & z e^{-z P_0} J_2   & e^{-z P_0}   & 0  & 0   & 0  & -z e^{-zP_0} P_3   & z e^{-zP_0} P_2 \\
  z e^{-zP_0} K_2   &  z e^{-zP_0} J_3  & 0  & -z e^{-zP_0} J_1  & 0  & e^{-zP_0} & 0 & z e^{-zP_0} P_3  & 0 &  -z e^{-zP_0} P_1 \\
  z e^{-zP_0} K_3   &  -z e^{-zP_0} J_2 &  z e^{-zP_0} J_1 & 0 & 0 & 0 &   e^{-zP_0}  &   -z e^{-zP_0} P_2 &  z e^{-zP_0} P_1 & 0  \\
  0 & 0 & 0 & 0 & 0 & 0 & 0  &  1 &  0  & 0 \\
  0 & 0 & 0 & 0 & 0 & 0 & 0  &  0 &  1  & 0 \\
  0 & 0 & 0 & 0 & 0 & 0 & 0  &  0 &  0  & 1           
\end{array}
\right)
\ee}

By computing the product (\ref{matrixp}) it is straightforward to prove that the group multiplication induces the following coproduct for the coordinate functions:
\begin{eqnarray}
\Delta_{G^\ast}\left( P_0\right) &=& P_0 \otimes 1 +1 \otimes P_0\nonumber \\
\Delta_{G^\ast}\left( P_i\right) &=& P_i \otimes e^{z P_0}+ 1 \otimes P_i \nonumber \\
\Delta_{G^\ast}\left( K_1\right) &=& K_1 \otimes e^{z P_0}+ 1 \otimes K_1 + z \left( J_2 \otimes P_3 - J_3 \otimes P_2 \right)\nonumber \\
\Delta_{G^\ast}\left( K_2\right) &=& K_2 \otimes e^{z P_0}+ 1 \otimes K_2 + z \left( J_3 \otimes P_1 - J_1 \otimes P_3 \right) \\
\Delta_{G^\ast}\left( K_3\right) &=& K_3 \otimes e^{z P_0}+ 1 \otimes K_3 +z \left( J_1 \otimes P_2- J_2 \otimes P_1 \right) \nonumber \\
\Delta_{G^\ast} \left( J_i \right) &=&  J_i \otimes 1 + 1 \otimes J_i .
\nonumber 
\end{eqnarray}
Now, in order to get the Poisson tensor that endows $G^\ast$ with the PL group structure associated with $(g^\ast,\delta^\ast)$ we have to follow the same computational steps as in the previous examples. In this case the group element and the coproduct map are not complicated, therefore the set of functions $\vec{F}$ has 22 elements:
\begin{eqnarray}
\vec{F}:=&& \left\{ 1, P_0,P_1,P_2,P_3,K_1,K_2,K_3,J_1,J_2,J_3,e^{z P_0},e^{-z P_0}, e^{-z P_0} P_1, e^{-z P_0} P_2,
e^{-z P_0} P_3, \right. \nonumber \\ 
&& \left. e^{-z P_0} K_1, e^{-z P_0} K_2, e^{-z P_0} K_3, e^{-z P_0} J_1, e^{-z P_0} J_2, e^{-z P_0} J_3  \right\}.
\label{functionsk}
\end{eqnarray}
The next step is to impose $\Delta_{G^\ast}$ to be a poisson morphism for the generic $Q_{ij}$ bracket (\ref{ansatz}). After solving the corresponding set of linear equations for the  11385 coefficients $\beta_{ijkl}$, we impose the linearized solution to be given by the structure tensor of the initial (3+1) Poincar\'e Lie algebra $g$. This leads to the following unique solution
\begin{eqnarray}
\left\{ P_0, P_i \right\}_\Lambda&=&0\nonumber \\
\left\{ P_0, K_{i} \right\}_\Lambda&=&-P_i \nonumber \\
\left\{ P_0, J_i \right\}_\Lambda&=& 0 \nonumber \\
\left\{ P_i, P_j \right\}_\Lambda&=&0\nonumber \\
\left\{ P_i, J_j \right\}_\Lambda&=& \epsilon_{ijk} P_k \label{kpois} \\
\left\{ K_i, K_j \right\}_\Lambda&=&-\epsilon_{ijk} J_k\nonumber \\
\left\{ K_i,J_j \right\}_\Lambda&=&\epsilon_{ijk} K_k\nonumber \\
\left\{ J_i,J_j \right\}_\Lambda&=&\epsilon_{ijk} J_k \nonumber\\ 
\left\{ P_i, K_j \right\}_\Lambda&=&\left( \frac{1}{2z} \left(1-e^{2z P_0} \right) +\frac{z}{2} \left(P_1^2+P_2^2+P_3^2 \right) \right)\delta_{ij}- z\,P_i\,P_j \nonumber
\end{eqnarray}
that fulfills the Jacobi identity and gives us the Poisson tensor $\Lambda$.

Note that, since all the $P_i$ coordinate functions Poisson-commute, the quantization of this Poisson-Hopf algebra is immediate and gives exactly the $\kappa$-Poincar\'e algebra in the bicrossproduct basis~\cite{Majidbicross}, where $z=1/\kappa$. Evidently, if we perform the change of coordinate functions given by
\be
P_i'=e^{-\frac{z}{2} P_0} P_i, \qquad K_i'=e^{-\frac{z}{2} P_0} K_i - \frac{z}{2} \epsilon_{ijk} J_j P_k
\label{cambiok}
\ee
then the new Poisson brackets read
\begin{eqnarray}
\left\{ P_0, P_i' \right\}_\Lambda&=&0\nonumber \\
\left\{ P_0, K_{i} \right\}_\Lambda&=&-P_i'\nonumber  \\
\left\{ P_0, J_i \right\}_\Lambda&=& 0 \nonumber \\
\left\{ P_i', P_j'\right\}_\Lambda&=&0\nonumber \\
\left\{ P_i', K_j' \right\}_\Lambda&=& -\delta_{ij} \frac{\sinh(z P_0)}{z} \label{poisk2}\\
\left\{ K_i', K_j' \right\}_\Lambda&=&-\epsilon_{ijk} \left( J_k \cosh(z P_0) - \frac{z^2}{4} P_k' (J_1 P_1'+J_2 P_2' + J_3 P_3') \right)\nonumber \\
\left\{ K_i',J_j \right\}_\Lambda&=&\epsilon_{ijk} K_k'\nonumber\\
\left\{ J_i,J_j \right\}_\Lambda&=&\epsilon_{ijk} J_k
\nonumber
\end{eqnarray}
and the coproduct maps are transformed into:
\begin{eqnarray}
\Delta_{G^\ast}\left( P_0\right) &=& P_0 \otimes 1 +1 \otimes P_0 \nonumber\\
\Delta_{G^\ast}\left( P_i\right) &=& P_i \otimes e^{\frac{z}{2} P_0}+ e^{-\frac{z}{2} P_0} \otimes P_i \nonumber\\
\Delta_{G^\ast}\left( K_i\right) &=& K_i \otimes e^{\frac{z}{2} P_0}+ e^{-\frac{z}{2} P_0} \otimes K_i + \frac{z}{2}\epsilon_{ijk}\left( P_j \otimes J_k e^{\frac{z}{2}P_0} + e^{-\frac{z}{2}P_0} J_j  \otimes P_k \right)\\
\Delta_{G^\ast} \left( J_i \right) &=&  J_i \otimes 1 + 1 \otimes J_i
\nonumber
\end{eqnarray}
The direct quantization of this Poisson-Hopf algebra is straightforward (there are no ordering ambiguities and, once again, the symmetry requirement from Remark 2 holds) and leads to the original presentation of the $\kappa$-Poincar\'e quantum algebra given in~\cite{LNR}.  


\section{Adding a twist to $\kappa$-Poincar\'e algebra}

As a final example, we will find the deformation of the Poncar\'e algebra that is obtained by twisting the previous $\kappa$-Poincar\'e deformation by a twist transformation~\cite{Drtwist} generated by the classical $r$-matrix 
\be
\tilde r=\eta\,(j_3\wedge p_0).
\ee
This means that we are going to consider the same basis for the Poincar\'e Lie algebra as in the previous section, but now 
the classical $r-$matrix that generates the initial Lie bialgebra $(g,\delta_t)$ will be
\be
r_t=z \left(k_1 \wedge p_1 + p_2 \wedge p_2+ k_3 \wedge p_3 \right)+\eta \, j_3 \wedge p_0 .
\label{twistedr}
\ee
Therefore, we are aiming to recover the Poncar\'e deformation whose first order is given by eq.~(51) in the classification~\cite{Zakrcmp}, and whose complete structure has been given in~\cite{Dasz} by twisting the $\kappa$-Poincar\'e Hopf algebra through the twist operator $F$ generated by $\tilde r$. In contradistinction with~\cite{Dasz}, we will construct the full twisted $\kappa$-Poincar\'e deformation in a unique step and starting from the corresponding $\delta$ defined by (\ref{twistedr}), which reads:
\begin{eqnarray}
\delta_t(p_0)&=&0 \nonumber\\
\delta_t(p_1)&=&z \, p_1 \wedge p_0 + \eta \, p_0 \wedge p_2  \nonumber\\
\delta_t(p_2)&=&z \,  p_2 \wedge p_0 + \eta \, p_1 \wedge p_0  \nonumber\\
\delta_t(p_3)&=&z \, p_3 \wedge p_0  \nonumber\\
\delta_t(k_1)&=&z \, \left(k_1 \wedge p_0 + p_2 \wedge j_3 + j_2 \wedge p_3 \right) + \eta \, \left(p_0 \wedge k_2 + j_3 \wedge p_1 \right) \\
\delta_t(k_2)&=&z \, \left( j_3 \wedge p_1 + p_3 \wedge j_1 + k_2 \wedge p_0 \right) + \eta \left(k_1 \wedge p_0 + j_3 \wedge p_2 \right)  \nonumber\\
\delta_t(k_3)&=&z \left(p_1 \wedge j_2 + j_1 \wedge p_2  + k_3 \wedge p_0 \right) + \eta \, j_3 \wedge p_3 \nonumber\\
\delta_t(j_1)&=& \eta \, p_0 \wedge j_2 \nonumber \\
\delta_t(j_2)&=& \eta \, j_1 \wedge p_0 \nonumber \\
\delta_t(j_3)&=&0.
\nonumber
\end{eqnarray}  
As a consequence, the dual Lie bialgebra $(g^\ast_t,\delta^\ast)$ (which is again a non-coboundary Lie bialgebra) will have the following commutation rules:
\be
\begin{array}{lll}
\left[ p^0,p^1 \right]= -z p^1 - \eta p^2 \quad & \left[ p^0,p^2 \right]= -z p^2 + \eta p^1  \quad & \left[ p^0,p^3 \right]= -z p^3 \\[2pt]
\left[ p^0,k^1 \right]= -z k^1 - \eta k^2 \quad & \left[ p^0,k^2 \right]= -z k^2 + \eta k^1  \quad & \left[ p^0,k^3 \right]= -z k^3 \\[2pt]
\left[ p^0,j^1 \right]= -\eta j^2  \quad & \left[ p^0,j^2 \right]= \eta j^1  \quad & \left[ p^0,j^3 \right]= 0 \\[2pt]
\left[ p^i,p^j \right]= 0  \quad & \left[ p^i,k^j \right]= 0  \quad & \left[ p^1,j^1 \right]= 0 \\[2pt]
\left[ p^1,j^2 \right]= z k^3  \quad & \left[ p^1,j^3 \right]= -z k^2 - \eta k^1  \quad & \left[ p^2,j^1 \right]= -z k^3 \\[2pt]
\left[ p^2,j^2 \right]= 0  \quad & \left[ p^2,j^3 \right]= z k_1 - \eta k_2  \quad & \left[ p^3,j^1 \right]= z k^2 \\[2pt]
\left[ p^3,j^2 \right]= -z k^1  \quad & \left[ p^3,j^3 \right]= - \eta k^3  \quad & \left[ k^i,k^j \right]= 0 \\[2pt]
\left[ k^i,j^j \right]= 0  \quad & \left[ j^i,j^j \right]= 0  \quad & 
\end{array}
\ee
Note that the introduction of the twist implies that the dual Lie algebra $g^\ast_t$ and its Lie group $G^\ast_t$ are different (in fact, much more complicated) from those previously obtained in the $\kappa$-Poincar\'e case. From these new commutation relations we obtain the adjoint representation $\rho$ of $G^\ast_t$ and we use it to
construct the corresponding Lie group element through
\begin{eqnarray}
&& \!\!\!\! \!\!\!\! \!\!\!\! \!\!\!\! G^\ast_t=\exp \left(P_0 \rho(p^0) \right) \exp \left( P_1 \rho(p^1) \right) \exp \left(P_2 \rho(p^2) \right)
\exp \left(P_3 \rho(p^3) \right)\cdot \nonumber \\
&& \!\!\!\! \!\!\!\! 
\cdot\exp \left( K_1 \rho(k^1) \right) \exp \left(K_2 \rho(k^2) \right) \left(K_3 \rho(k^3) \right) \exp \left( J_1 \rho(j^1) \right) \exp \left(J_2 \rho(j^2) \right)
\exp \left(J_3 \rho(j^3) \right)
\label{gtwist}
\end{eqnarray}
where again $\{P_0,P_1,P_2,P_3,K_1,K_2,K_3,J_1,J_2,J_3 \}$ are the coordinate functions on the new $G^\ast_t$. The functional equations arising from the group multiplication can be again solved, and we get the following twisted coproduct for the coordinate functions on $G^\ast_t$
\begin{eqnarray}
\Delta_{G^\ast}\left( P_0\right) &=& P_0 \otimes 1 +1 \otimes P_0 \nonumber\\
\Delta_{G^\ast}\left( P_1\right) &=& P_1 \otimes e^{z P_0} \cos(\eta P_0) - P_2 \otimes e^{z P_0} \sin(\eta P_0)  + 1 \otimes P_1 \nonumber  \\
\Delta_{G^\ast}\left( P_2\right) &=& P_2 \otimes e^{z P_0} \cos(\eta P_0) + P_1 \otimes e^{z P_0} \sin(\eta P_0)  + 1 \otimes P_2 \nonumber \\
\Delta_{G^\ast}\left( P_3\right) &=& P_3 \otimes e^{z P_0}  + 1 \otimes P_3 \nonumber \\
\Delta_{G^\ast}\left( K_1\right) &=& K_1 \otimes e^{z P_0} \cos(\eta P_0 ) + 1 \otimes K_1 + z \left( J_2 \otimes P_3 \cos( \eta P_0) +J_1 \otimes  P_3 \sin( \eta P_0) 
-J_3 \otimes P_2 \right) -\nonumber\\
&& + \eta J_3 \otimes P_1 - K_2 \otimes e^{z P_0} \sin( \eta P_0)  \nonumber\\
\Delta_{G^\ast}\left( K_2\right) &=& K_2 \otimes e^{z P_0} \cos(\eta P_0 ) + 1 \otimes K_2 + z \left( J_3 \otimes P_1 - J_1 \otimes P_3 \cos(\eta P_0 ) + J_2
\otimes P_3 \sin(\eta P_0 ) \right)+ \nonumber \\
&& + \eta J_3 \otimes P_2 + K_1 \otimes e^{z P_0} \sin(\eta P_0 ) \label{cotwist}\\
\Delta_{G^\ast}\left( K_3\right) &=& K_3 \otimes e^{z P_0}+ 1 \otimes K_3 +z ( J_1 \otimes ( P_2 \cos(\eta P_0)  - P_1 \sin(\eta P_0) ) - \nonumber\\
&& - J_2 \otimes (P_1 \cos(\eta P_0) + P_2 \sin (\eta P_0) )  )+ \eta J_3 \otimes P_3 \nonumber\\
\Delta_{G^\ast} \left( J_1 \right) &=&  J_1 \otimes \cos(\eta P_0 )  - J_2 \otimes \sin (\eta P_0 ) + 1 \otimes J_1 \nonumber\\
\Delta_{G^\ast} \left( J_2 \right) &=&  J_2 \otimes \cos(\eta P_0 ) +  J_1 \otimes \sin (\eta P_0 ) + 1 \otimes J_2 \nonumber\\
\Delta_{G^\ast} \left( J_3 \right) &=&  J_3 \otimes 1 + 1 \otimes J_3
\nonumber
\end{eqnarray}
which has indeed new $\eta$-contributions coming from the twist, whose existence induces a much more complicated multiplication law on the dual group.  We remark that these expressions for the coproduct are different (and simpler) than the ones presented in~\cite{Dasz}, that would correspond to a different choice of the ordering of the exponentials when constructing the group element $G^\ast_t$ (\ref{gtwist}).

Now, in order to get the Poisson tensor, we would have to reproduce again the same algorithm as in the previous sections. In this case the set $\vec F$ containing all the functions appearing in the group element $G^\ast_t$ and  in the previous coproduct (\ref{cotwist}) would be much larger than in the $\kappa$-Poincar\'e case. Nevertheless, since we expect that the twist will not affect the commutation rules (therefore, the PL bracket should be the same), and since we know that the final solution for the full dual PL bracket is unique, we firstly tried with the `reduced' $\kappa$-Poincar\'e set of functions $\vec F$ given in (\ref{functionsk}), thus expecting that it would provide the solution. This assumption worked, since 
after obtaining the most general PL bracket compatible with the twisted coproduct (\ref{cotwist}) and constructed in terms of this restricted set of functions, we imposed on it the linearization condition and the unique final solution so obtained was just the same (\ref{kpois}) as in the non-twisted $\kappa$-deformation.

Finally, as far as the quantization is concerned we remark that the twist do not generate further ordering problems in the coproduct, and the Poisson-Hopf algebra can be straightforwardly quantized, thus  giving rise to a twisted $\kappa$-Poincar\'e algebra in a bicrossproduct basis. Evidently, if we perform the same change of coordinate functions (\ref{cambiok}) 
then the new Poisson brackets are again (\ref{poisk2}) and the twisted coproducts read
\begin{eqnarray}
\Delta_{G^\ast}\left( P_0\right) &=& P_0 \otimes 1 +1 \otimes P_0 \nonumber\\
\Delta_{G^\ast}\left( P_1'\right) &=& P_1' \otimes e^{\frac{z}{2} P_0} \cos(\eta P_0) + e^{-\frac{z}{2} P_0} \otimes P_1'- P_2' \otimes e^{\frac{z}{2} P_0} \sin(\eta P_0) \nonumber\\
\Delta_{G^\ast}\left( P_2'\right) &=& P_2' \otimes e^{\frac{z}{2} P_0} \cos(\eta P_0) + e^{-\frac{z}{2} P_0} \otimes P_2'+ P_1' \otimes e^{\frac{z}{2} P_0} \sin(\eta P_0 \nonumber)\\
\Delta_{G^\ast}\left( P_3'\right) &=& P_3' \otimes e^{\frac{z}{2} P_0}+ e^{-\frac{z}{2} P_0} \otimes P_3' \nonumber\\
\Delta_{G^\ast}\left( K_1'\right) &=& K_1' \otimes e^{\frac{z}{2} P_0} \cos(\eta P_0)+ e^{-\frac{z}{2} P_0} \otimes K_1' + \frac{z}{2} 
\left( P_2' \otimes J_3 e^{\frac{z}{2}P_0} \cos(\eta P_0) + P_1' \otimes J_3 e^{\frac{z}{2}P_0} \sin(\eta P_0)   + \right. \nonumber\\
&& \left. + e^{-\frac{z}{2}P_0} J_1  \otimes P_3' \sin(\eta P_0) + e^{-\frac{z}{2}P_0} J_2  \otimes P_3' \cos(\eta P_0)-P_3' \otimes J_2 e^{\frac{z}{2}P_0} - J_3 e^{-\frac{z}{2}P_0} \otimes P_2' \right) - \nonumber\\
&& -K_2' \otimes e^{\frac{z}{2}P_0} \sin(\eta P_0)+ \eta J_3 e^{-\frac{z}{2}P_0} \otimes P_1'\nonumber\\
\Delta_{G^\ast}\left( K_2'\right) &=& K_2' \otimes e^{\frac{z}{2} P_0} \cos(\eta P_0)+ e^{-\frac{z}{2} P_0} \otimes K_2' + \frac{z}{2} 
\left( P_2' \otimes J_3 e^{\frac{z}{2}P_0} \sin(\eta P_0) - P_1' \otimes J_3 e^{\frac{z}{2}P_0} \cos(\eta P_0)   + \right. \nonumber\\
&& \left. + e^{-\frac{z}{2}P_0} J_2  \otimes P_3' \sin(\eta P_0) -e^{-\frac{z}{2}P_0} J_1  \otimes P_3' \cos(\eta P_0)
+P_3' \otimes J_1 e^{\frac{z}{2}P_0} + J_3 e^{-\frac{z}{2}P_0} \otimes P_1' \right) - \nonumber\\
&& -K_1' \otimes e^{\frac{z}{2}P_0} \sin(\eta P_0)+ \eta J_3 e^{-\frac{z}{2}P_0} \otimes P_2'\\
\Delta_{G^\ast}\left( K_3'\right) &=& K_3' \otimes e^{\frac{z}{2} P_0}+ e^{-\frac{z}{2} P_0} \otimes K_3' 
+ \frac{z}{2} \left( -P_2' \otimes J_2 e^{\frac{z}{2}P_0} \sin(\eta P_0) - P_1' \otimes J_2 e^{\frac{z}{2}P_0} \cos(\eta P_0)   + \right. \nonumber\\
&&  - e^{-\frac{z}{2}P_0} J_1  \otimes P_1' \sin(\eta P_0) -e^{-\frac{z}{2}P_0} J_2  \otimes P_1' \cos(\eta P_0)
-P_1' \otimes J_1 e^{\frac{z}{2}P_0} \sin(\eta P_0) + \nonumber\\
&&+ \left.  J_1 e^{-\frac{z}{2}P_0} \otimes P_2' \cos(\eta P_0) -P_2' \otimes J_1 e^{\frac{z}{2}P_0} \cos(\eta P_0) 
-J_2 e^{-\frac{z}{2}P_0} \otimes P_2'  \sin(\eta P_0)\right) + \nonumber\\
&& +\eta J_3 e^{-\frac{z}{2}P_0} \otimes P_3' \nonumber\\
\Delta_{G^\ast}\left( J_1 \right) &=&  J_1 \otimes \cos(\eta P_0) - J_2 \otimes \sin(\eta P_0) + 1 \otimes J_1 \nonumber\\
\Delta_{G^\ast}\left( J_2 \right) &=&  J_2 \otimes \cos(\eta P_0) + J_1 \otimes \sin(\eta P_0) + 1 \otimes J_2  \nonumber\\
\Delta_{G^\ast}\left( J_3 \right) &=&  J_3 \otimes 1 + 1 \otimes J_3.
\nonumber
\end{eqnarray}
These expressions give the classical analogue of the twisted $\kappa$-Poincar\'e algebra in the basis (\ref{cambiok}). Again, the quantization of this Poisson-Hopf algebra is straightforward, and the usual $\kappa$-Poincar\'e algebra is obtained in the limit $\eta\to 0$. Note that, by following the method here presented, the compatibility between this coproduct and the commutation rules is the only non-commutative computation that has to be performed in the full process of the construction of this (quite complicated) quantum algebra. Finally, we remark that the latter coproduct does indeed fulfil the `quantum symmetry property' stated in Remark 2 with respect to the deformation parameter $z$, but a further change of coordinates would be needed in order to get that the $\eta$-terms coming from the twist would present the same symmetry. Nevertheless, such terms induced by the twist do not generate any ordering ambiguity as far as the quantization is concerned, and the latter change of coordinates can be avoided.


\section{Concluding remarks}

The main objective of this paper has been to show that the explicit construction of a quantum algebra (starting from their defining cocommutator map $\delta$) can be systematically obtained by considering the quantum algebra as a quantum dual group, {\em i.e.}, as the quantization of the unique Poisson Lie structure on the dual group $G^\ast$ that is determined by the dual Lie bialgebra $(g^\ast,\delta^\ast)$. Although all the theoretical foundations of this approach (namely, the `quantum duality principle') are indeed well known, we think that the algorithmic implementation  we propose can provide a helpful tool in order to solve explicitly  the Lie bialgebra quantization problem in many new interesting cases.

In fact, the main advantage of this approach is that, for most of the cases, the computation of such PL structure on $G^\ast$ can be fully performed with the aid of a symbolic computation program, and only the final stage of the quantization of this Poisson-Hopf algebra requires the use of noncommutative variables. Moreover, this latter step can be strongly simplified by finding the appropriate set of local coordinates on $G^\ast$ that minimize the ordering ambiguities prior to quantization, and this is again a `classical' Poisson algebra problem. The same would happen with the search of the Casimir functions for the Poisson structure $\Lambda$, whose quantization would provide the Casimir operators for the corresponding quantum algebra. This methodology has been illustrated through several examples in different dimensions, including the construction of a (quite complicated) twisted (3+1) $\kappa$-Poincar\'e algebra.

We think that this approach is interesting both from a technical viewpoint and also from a more fundamental one, since it provides a more clear interpretation of several quantum algebra features by making use of a purely group theoretical framework. For instance, we have seen that if we perform a twist deformation on a given quantum algebra whose dual Lie algebra is $g^\ast$, the twist generates new terms in the commutation rules of the new twisted dual algebra $g_t^\ast$, which becomes `less commutative'. Therefore, the coproduct coming from the new twisted group law for $G_t^\ast$ has new terms ({\em i.e.} it is `less cocommutative'). Nevertheless, the Poisson brackets on $G_t^\ast$ are not modified by the twist with respect to the ones in the original $G^\ast$, which means that the twist is connecting two different dual groups that share a common PL bracket. 

Also, this method for the construction of quantum algebras emphasizes the fact that the semiclassical limit of a quantum algebra is just given by the dual PL group $(G^\ast, \Delta_{G^\ast}, \Lambda)$. As a consequence, the corresponding PL dynamics on such dual group will be the semiclassical footprint of the full quantum algebra dynamics. In this respect,  the quest for the physical significance of the coproduct as the algebraic way to define observables for composite systems could be also enlighted by interpreting coproducts as dual group multiplications. Moreover, the way to get systematically a parametrization of the group manifold of $G^\ast$ in terms of coordinates such that the corresponding PL structure is free from ordering ambiguities, remains as an open problem.

Finally, the `bottom-up' approach here presented could be very useful in order to construct new complicated quantum algebras whose first order deformations are deduced from physical considerations. This would be the case for quantum deformations of the (2+1) de Sitter and Anti de Sitter Lie algebras whose classical $r$-matrices have recently arised from a Drinfel'd double structure which is consistent with the pairing structure of the Lie algebras of isometries for the solutions of (2+1) gravity models~\cite{BHMplb}. Also, in order to obtain Double Special Relativity models with non-vanishing cosmological constant~\cite{Starodutseb, cm1} the obtention of the Drinfel'd-Jimbo deformation for $so(3,2)$  and $so(4,1)$ by following the present approach would be meaningful, since  the deformation would be constructed from the very beginning in the physically relevant kinematical basis. In this way the problem of the definition of the physical generators out of $q$-Serre relations is avoided, as well as the need of using real forms is (see, for instance,~\cite{LRNreal,Dobrev}). Work on all these lines is in progress and will be presented elsewhere.


\section*{Acknowledgements}

This work was partially supported by the Spanish MINECO   under grant   MTM2010-18556 (with EU-FEDER support)
and by INFN--MINECO (grant AIC-D-2011-0711). The authors acknowledge all the referees for their constructive and useful comments.




\begin{thebibliography}{99}



\bibitem{Dri}
Drinfel'd V G 1987  {\it Quantum Groups}, in:
  Gleason A V (Ed.),   {\it Proc. Int. Cong. Math.
Berkeley 1986},  (Providence: AMS) p. 798

\bibitem {FRT} 
  Reshetikhin N Y,  Takhtadzhyan L A and Faddeev L D 1990  
{\it Leningrad Math. J.} {\bf 1}  193   

\bibitem{fuchs}  Fuchs J 1992 
 {\em Affine Lie Algebras and Quantum Groups}
(Cambridge: Cambridge University Press)

 \bibitem{schnider}   Schnider S, Sternberg S 1993
 {\em  Quantum Groups: from Coalgebras to Drinfeld Algebras}
(Cambridge MA: International Press)

\bibitem{charipressley}  Chari V, Pressley A 1995 {\it A Guide to Quantum
Groups}
(Cambridge: Cambridge University Press)

\bibitem{majid}  Majid S 1995 {\it Foundations of Quantum
Group Theory}  (Cambridge: Cambridge University Press)

\bibitem{DriPL} Drinfel'd V G 1983 {\it Soviet Math. Dokl}. {\bf 27} 68

\bibitem{Semquantum} Semenov-Tian-Shanskii M A 1992 {\it Theor. Math. Phys.} {\bf 93} 302 

\bibitem{Bidegain}  Bidegain F, Pinczon G 1996 {\it Commun. Math. Phys.} {\bf 179} 295

 \bibitem{gomez} Gomez X 2000 {\it J. Math. Phys}. {\bf 41} 
4939

\bibitem{BBM3d} Ballesteros A,  Blasco A, Musso F 2012
{\it J. Phys. A: Math. Theor.} {\bf 45} 175204

\bibitem{LV} Ballesteros A, Blasco A, Musso F 2011 {\it Phys. Lett. A} {\bf{375}} 3370

\bibitem{EK} Etingof P, Kazhdan D 1996 {\it Selecta Math.} {\bf 2}  1

\bibitem{EH} Enriquez B, Halbout G 2010 {\it Ann. Math.} {\bf 171}  1267

\bibitem{Drtwist}
Drinfeld V G 1990
{\it Leningrad Math. J.} {\bf 1} 1419

\bibitem{Reshetikhin}
  Reshetikhin N 1990
{\it Lett. Math. Phys.} {\bf 20} 331

\bibitem{Mudrov}
  Mudrov A I 1998
{\it J. Phys. A: Math. Gen.} {\bf 31} 6219

\bibitem{twistedP} Lukierski J, Ruegg H, Tolstoy V N, Nowicki A 1994
{\it J. Phys. A: Math. Gen.} {\bf 27} 2389

\bibitem{KLM}  Kulish P P, Lyakhovsky V D,  Mudrov A I 1999  {\it J.
Math. Phys} {\bf 40} 4569 

\bibitem{LukiWoroPLB} Lukierski L, Woronowicz M 2006  {\it Phys. Lett. B} {\bf{633}} 116

\bibitem{Dasz}  Daszkiewicz M  2008  {\it Int. J.
Mod. Phys A} {\bf 23} 4387 

\bibitem{Tjin} Tjin T 1992 {\it Int. J. Mod. Phys. A} {\bf 7} 6175

\bibitem{ballesteros04} Ballesteros A, Celeghini E, del Olmo M A 2004
  {\it J. Phys. A: Math. Gen.} {\bf 37} 4231

\bibitem{STS} Semenov-Tyan-Shanskii MA 1992 {\it Theor. Math. Phys.} {\bf 93}  1292

\bibitem{Gav} Gavarini F 2002 {\it Ann. Inst. Fourier} {\bf 52}  809

\bibitem{GC} Ciccoli N, Gavarini F 2006 {\it Adv. Math.} {\bf 199}  104

\bibitem{BBMpullback} Musso F, Ballesteros A,  Blasco A  2012
{\it AIP Conf. Proc.} {\bf 1460} 211

\bibitem{BHOSpl} Ballesteros A, Herranz F J, del Olmo M A, Santander M 1995  {\it J.
Math. Phys} {\bf 36} 631 

\bibitem{LM}  
Lyakhovsky V, Mudrov A I 1992
{\it J. Phys. A: Math. Gen.} {\bf 25} L1139

\bibitem{MudrovPet}  Mudrov A I 1994
{\it Vest. St. Petersburg Univ.} {\bf 4} 3

\bibitem{MudrovJMP}  Mudrov A I 1997
{\it J. Math. Phys}.  {\bf 38}  476   

\bibitem{Jimbo}  Jimbo M 1985
{\it Lett. Math. Phys}.  {\bf 10}  63   

\bibitem{LNRT} Lukierski J, Nowicki A, Ruegg H, Tolstoy V N 1991  {\it Phys. Lett. B} {\bf{264}} 331

\bibitem{LNR} Lukierski J, Nowicki A, Ruegg H 1992  {\it Phys. Lett. B} {\bf{293}} 344

\bibitem{euclideo} Sobczyk J 1996 {\it J. Phys. A: Math. Gen.} {\bf 29}  2887
 
\bibitem{galileo} Kowalczyk E 1997 {\it Acta Phys. Pol. B} {\bf 28}  1893

\bibitem{anna} Opanowicz A 1998 {\it J. Phys. A: Math. Gen.} {\bf 31}  8387

 \bibitem{brihaye} Brihaye Y, Kowalczyk E, Maslanka P 2001 
{\it Mod. Phys. Lett. A} {\bf 16} 321

\bibitem{checos} Vysok\'y J 2011 {\it Poisson structures on Lie groups} Dipl. Th. (Prague: CTU) 

\bibitem{Marmo} Grabowski J, Marmo G, Michor P W 1999  {\it Mod. Phys. Lett. A} {\bf{14}} 2109

\bibitem{beyond} Ballesteros A,  Herranz F J, del Olmo M A, Santander M 1995
{\it J. Phys. A: Math. Gen.} {\bf 28} 941

\bibitem{Majidbicross} Majid S, Ruegg H 1994  {\it Phys. Lett. B} {\bf{334}} 348

\bibitem{Zakrcmp} Zakrzewski S 1997
{\it Commun. Math. Phys.} {\bf 185} 285

\bibitem{BHMplb} Ballesteros A, Herranz F J, Meusburger C 2010 {\it Phys. Lett. B}  {\bf 687} 375

\bibitem{Starodutseb} Amelino-Camelia G, Smolin L, Starodubtsev A 2004 {\it Class. Quantum Grav.}
 {\bf{21}} 3095
 
 \bibitem{cm1}  Meusburger C, Schroers B J 2008
 {\it J. Math. Phys.} {\bf{49}} 083510

\bibitem{LRNreal} Lukierski J, Ruegg H, Nowicki A  1991  {\it Phys. Lett. B} {\bf{271}} 321

\bibitem{Dobrev} Dobrev V K 1993 {\it J. Phys. A: Math. Gen.} {\bf 26}  1317











\end{thebibliography}
\end{document}